\newif\ifpreprint

\preprintfalse % Enable for double column reprint

\ifpreprint
\documentclass[aip,jcp,amsmath,amssymb,preprint]{revtex4-1}
\else
\documentclass[aip,jcp,amsmath,amssymb,reprint]{revtex4-1}
\fi

\usepackage{amsmath}
\usepackage{amssymb}
\usepackage{graphicx}
\usepackage{color}
\usepackage{braket}
\usepackage{xspace}
\usepackage{enumerate}
\usepackage[note-name=,use-sort-key=false]{notes2bib}

\newcommand*{\kcal}{kcal mol$^{-1}$\xspace}

\newcommand*{\Eh}{$E_{\rm h}$\xspace}

\begin{document}

\title{An adaptive configuration interaction approach for strongly correlated electrons with tunable accuracy}
\author{Jeffrey B. Schriber and Francesco A. Evangelista}
\affiliation{Department of Chemistry and Cherry L. Emerson Center for Scientific Computation, Emory University, Atlanta, Georgia, 30322, USA}

\begin{abstract}
We introduce a new procedure for iterative selection of determinant spaces capable of describing highly correlated systems. 
This adaptive configuration interaction (ACI) determines an optimal basis by an iterative procedure in which the determinant space is expanded and coarse grained until self consistency.
Two importance criteria control the selection process and tune the ACI to a user-defined level of accuracy.
The ACI is shown to yield potential energy curves of N$_2$ with nearly constant errors, and it predicts singlet-triplet splittings of acenes up to decacene that are in good agreement with the density matrix renormalization group.
\end{abstract}

\maketitle

Most popular methods in electronic structure theory by some means attempt to exploit the sparsity of full configuration interaction (FCI)
wave functions.\cite{sherrillCIrev} The exponential scaling of the number of determinants with respect to the number of orbitals required for FCI calculations
prevents its use for all but trivially small systems, or for active space calculations no larger than 18 electrons in 18 orbitals.
Recently, wave function factorization techniques such as the density matrix renormalization group,
\cite{white1992density,chan2002highly,moritz2005relativistic, kurashige2009high, olivares2015ab}  and stochastic CI approaches such as Monte Carlo CI (MCCI)\cite{greer1995estimating, kelly2014monte,
coe2012development, coe2014applying} and FCI Quantum Monte Carlo (FCIQMC)\cite{booth2009fermion,cleland2010communications,Petruzielo:2012ha,Tenno:2013kd} have risen as promising alternatives to FCI and complete active space CI (CASCI), allowing for the description of chemically interesting systems.\cite{kurashige2013entangled,Booth:2013er}

In this study, we propose a new adaptive configuration interaction (ACI) method that produces compact wave functions with tunable accuracy.
The ACI is based on the framework of selected CI,\cite{bender1969studies,huron1973iterative,buenker1974individualized,harrison1991approximating}
which recently has received renewed attention.\cite{garcia1995iterative,neese2003spectroscopy,nakatsuji2005iterative,abrams2005important,
bytautas2009priori,roth2009importance,evangelista2014adaptive,knowles2015compressive,liu2016ici,Tubman:2016wi} 
It uses two parameters to control the treatment of electron correlation.
As will be shown, a remarkable property of the ACI is its ability to compute electronic energies with almost perfect control over the energy error.
Additionally, we demonstrate that the ACI is a viable alternative to traditional 
%methods that describe strongly correlated electrons via a
 complete active space (CAS) methods by performing ACI computations on active spaces that are outside the reach of CASCI.

\begin{figure*}[ht!]
\ifpreprint
	\includegraphics[width=6.5in]{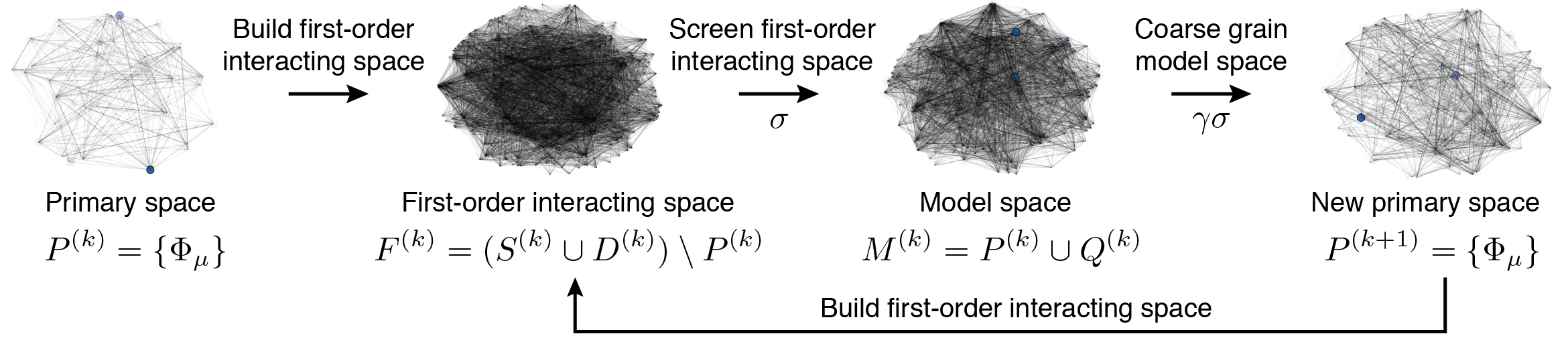}
\else
	\includegraphics[width=6.5in]{figure_1_comm.png}
\fi
\caption{Evolution of determinant spaces in the ACI algorithm. Each node represents a determinant, and the edges represent coupling through the Hamiltonian between two nodes. The
edges are weighted by the magnitude of this coupling, and the nodes are weighted proportionally to the square modulus of the determinant coefficient ($|C_{\mu}|^2$).}
\label{fig:aci_algorithm}
\end{figure*}

Given a set of orthonormalized one-electron molecular orbitals, $\{ \phi_p \}$, the ACI requires the user to specify the number of electrons, the spin multiplicity, and two orbital subsets: doubly occupied orbitals and active orbitals.  The latter are partially occupied in all determinants generated by the ACI.
The ACI procedure is illustrated in Fig.~\ref{fig:aci_algorithm} and consists of the following steps: 
\begin{enumerate}[i)]
\item 
At each iteration $k$ we define the space of reference determinants [$P^{(k)}$]:
\begin{equation}
	P^{(k)} = \{\Phi_{\mu}~: \mu = 1, 2, \dots, d_k\},
\end{equation}
where $d_k$ is the dimension of the $P^{(k)}$ space.  To this space, we associate the configuration interaction wave function $\Psi_P^{(k)}$, defined as:
\begin{equation}
\ket{\Psi_P^{(k)}} = \sum_{\mu = 1}^{d_k} C_\mu \ket{\Phi_{\mu}},
\end{equation}
where the coefficients $C_\mu$ are determined by diagonalizing the Hamiltonian in the space $P^{(k)}$.
In most cases, we begin the ACI process with an initial reference space, $P^{(0)}$, that contains a single determinant, though a set of determinants can be used to speed convergence. 
\item
From the reference space $P^{(k)}$, all singly and doubly excited determinants are generated.
For a given $P^{(k)}$ space, we define the usual first-order
interacting space (FOIS), $F^{(k)}$, as the union of all unique singly [$S^{(k)}$] and doubly [$D^{(k)}$] excited determinants out of the reference space:
\begin{equation}
	F^{(k)}  = (S^{(k)} \cup D^{(k)}) \setminus P^{(k)}.
\end{equation}
Denoting the occupied (virtual) orbitals of determinant $\Phi_{\mu} \in P^{(k)}$ as $i, j, \dots$ ($a, b, \dots$), then $S^{(k)}$ and $D^{(k)}$ may be written compactly as  $S^{(k)} = \{\hat{a}_{a}^{\dagger}\hat{a}_{i}\Phi_{\mu} ~: \forall \Phi_\mu \in P^{(k)}\}$ and $D^{(k)} = \{\hat{a}_{a}^{\dagger}\hat{a}_{b}^{\dagger}\hat{a}_{j}\hat{a}_{i}\Phi_{\mu} ~: \forall\Phi_\mu \in P^{(k)}\}$.

\item
To each determinant in $F^{(k)}$ we associate an estimate of the energy contribution.
Following degeneracy-corrected perturbation theory,\cite{assfeld1995degeneracy} we consider the two-by-two Hamiltonian for a determinant $\Phi_{I} \in F^{(k)}$ interacting with a the $P$-space wave function $\Psi_P^{(k)}$:
\begin{equation}
\mathbf{H} = 
\begin{pmatrix}
\bra{\Psi_P^{(k)}} \hat{H} \ket{\Psi_P^{(k)}} & \bra{\Psi_P^{(k)}} \hat{H} \ket{\Phi_I}  \\
\bra{\Phi_I} \hat{H} \ket{\Psi_P^{(k)}} & \bra{\Phi_I} \hat{H} \ket{\Phi_I} 
\end{pmatrix}
=
\begin{pmatrix}
E_{P} & V \\
V^* & E_{I}
\end{pmatrix}.
\end{equation}
Diagonalization of $\mathbf{H}$ yields two real eigenvalues ($\lambda_1,\lambda_2$, $\lambda_1 \leq \lambda_2$), the lower of which differs from the energy of $\Psi_P^{(k)}$ ($E_P$) by:
\begin{equation} \label{eq:importance}
\epsilon(\Phi_I) = \lambda_1 - E_P  = \frac{\Delta}{2} - \sqrt{\frac{\Delta^{2}}{4} + |V|^{2}},
\end{equation}
where $\Delta = E_I - E_P$.
Eq.~\eqref{eq:importance} defines the energy importance criterion used in ACI to screen the first-order interacting space.

\item 
Using the energy importance criterion we define the \textit{secondary space} $Q^{(k)}$, the set of the most important determinants in $F^{(k)}$.
To build $Q^{(k)}$, we use an aimed selection scheme.\cite{angeli1997multireference2}  Firstly, we sort the set $F^{(k)}$ in decreasing order according to $|\epsilon(\Phi_I)|$, the absolute value of the energy importance criterion.
Secondly, starting from the determinant with the lowest $|\epsilon(\Phi_I)|$, we exclude all those elements of $F^{(k)}$ such that the cumulative energy error is less than a user-specified threshold $\sigma$ expressed in units of m\Eh:
\begin{equation}
\sum_{\Phi_I \in F^{(k)} \setminus Q^{(k)}} |\epsilon(\Phi_{I})| \leq \sigma.
\end{equation}
The determinants that are not discarded from $F^{(k)}$ form the set $Q^{(k)}$.

\item 
With the $Q^{(k)}$ space built, we can define the \emph{total model space} at iteration $k$ [$M^{(k)}$] as the union between the reference space and the secondary space:
\begin{equation}
	M^{(k)} = P^{(k)} \cup Q^{(k)},
\end{equation}
and diagonalize the Hamiltonian in the space $M^{(k)}$ to obtain the model space wave function:
\begin{equation}
\ket{\Psi_M^{(k)}} = \sum_{\Phi_{I} \in M^{(k)}}  C_{I} \ket{\Phi_{I}},
\end{equation}  
and the associated energy $E_M^{(k)}$.
The model space energy may be corrected for the contributions of the determinants excluded from the secondary space [$\Phi_I \in F^{(k)} \setminus Q^{(k)}$] using the second-order perturbative estimate:
\begin{equation} \label{eq:pt2}
E_F^{(k)} \approx E_M^{(k)} + \sum_{\Phi_I \in F^{(k)} \setminus Q^{(k)}} \epsilon(\Phi_{I}).
\end{equation}

\item
Rather than directly augmenting the total model space as the iterations proceed, as is traditionally done in selected
CI methods, we coarse grain the space $M^{(k)}$ to form an updated reference space $P^{(k+1)}$. Specifically, the $M^{(k)}$-space determinants are sorted according to the square of the CI coefficients ($|C_{I}|^2$) in decreasing order.
Determinants are progressively included in $P^{(k+1)}$ until the sum of the squared coefficients is less than $1-\gamma\sigma$, where $\gamma$ is a constant that has units of (energy)$^{-1}$:
\begin{equation}
\sum_{\Phi_\mu \in P^{(k+1)}} |C_\mu|^2 < 1 -\gamma\sigma.
\end{equation}

\item
Steps i--vi are repeated until the energy of the $M^{(k)}$ space is converged. This convergence of the energy coincides with 
the convergence of $P^{(k)}$ and $M^{(k)}$ with respect to the determinants included.
\end{enumerate}  

ACI improves upon previous selected CI methods like
CIPSI\cite{huron1973iterative} and CI+PT\cite{harrison1991approximating} in a number
of important ways. Firstly, the aimed selection procedure gives the user \textit{a priori} control over the absolute error in a computation. Additionally, 
the coarse-graining step (vi) increases the efficiency of the selection process (analogous to the initiator approximation of FCIQMC)\cite{cleland2010communications} and decreases the dependence on the starting wave function guess.

For all ACI calculations, the parameters $\sigma$ and $\gamma$ are directly related to the desired energy accuracy.
We found it convenient to assume a constant value of $\gamma$, and in this work all results were obtained using $\gamma$ = 1 m$E_{\rm h}^{-1}$.
Accordingly, ACI results will be denoted as ACI($\sigma$), while the ACI energy corrected for the determinant excluded from the secondary space [Eq.~\eqref{eq:pt2}] will instead be indicated as ACI($\sigma$)+PT2.

\begin{figure}[t!]
\centering
    \includegraphics{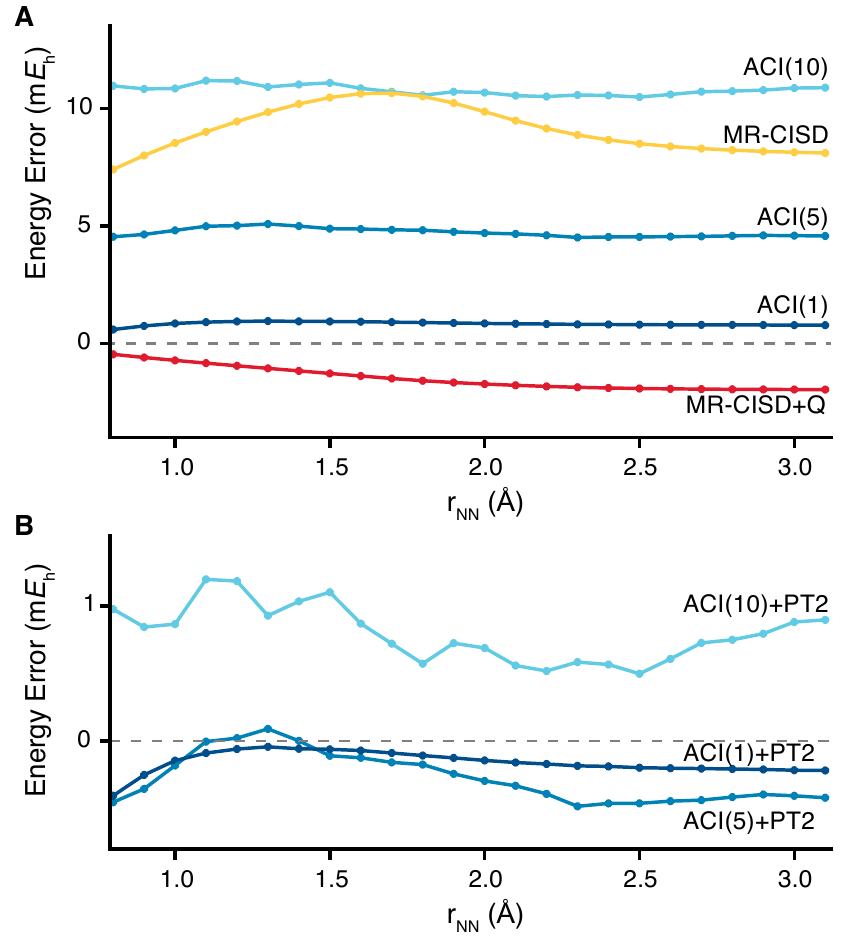}
    \caption{Ground-state potential energy curve of N$_2$ computed with the cc-pVDZ basis set.  (A) Energy errors with respect to FCI for the ACI, MR-CISD, and MR-CISD+Q.  (B) Energy errors with respect to FCI for the ACI plus the second-order energy correction [Eq.~\eqref{eq:pt2}].  ACI results used restricted Hartree--Fock orbitals while MR-CISD and MR-CISD+Q results are based on a CASSCF(6,6) reference.
The 1s-like orbitals of nitrogen were frozen in all correlated computations. }
    \label{fig:n2_comp}
\end{figure}

To illustrate the ability of ACI to determine molecular energies with nearly constant accuracy along a potential energy surface, we examine the dissociation of N$_2$.\bibnote{
To maximize efficiency, ACI works in the basis of Slater determinants rather than configuration state functions.  Consequently, $P^{(k)}$ and $M^{(k)}$ may not form spin complete sets.  To bypass this issue, in certain cases we have enforced spin completeness by appropriate augmenting $P^{(k)}$ and $M^{(k)}$.
In practice, correcting for spin incompleteness is only necessary to describe near-degenerate states of different spin.  Therefore, in this work this procedure is only applied to our N$_2$ computations to recover the correct asymptotic dissociation limit.
}
Figure~\ref{fig:n2_comp} shows the error with respect to FCI for the ground-state potential energy curve of N$_2$ computed with ACI using canonical restricted Hartree--Fock (RHF) orbitals.
In addition, we plot results for the internally-contracted multireference CISD (MR-CISD),\cite{werner1988efficient} and MR-CISD with Davidson's correction (MR-CISD+Q)\cite{langhoff1974configuration} based on a CAS self-consistent-field reference with six electrons in six orbitals [CASSCF(6,6)]. MR-CISD and MR-CISD+Q data from Ref \citenum{li2016towards} was used.
Figure~\ref{fig:n2_comp}A illustrates a distinguishing factor of the ACI: the absolute error at each point along the curve is reliably estimated by the energy threshold $\sigma$.
Moreover, while the ACI(10) curve displays noticeable microscopic discontinuity, the ACI(5) and ACI(1) curves are progressively smoother.
The inclusion of the second-order perturbative correction (see Fig.~\ref{fig:n2_comp}B) leads to curves that are approximately within 1 m\Eh from the FCI energy.
In comparison, MR-CISD shows fairly constant error throughout the dissociation, but with a noticeable increase near 1.6 \AA{}. With the $+Q$ correction,
the error is fairly constant across the potential, though with a slight decrease in accuracy towards dissociation. Additionally, these energies are not variational.

\begin{table}[t!]
\centering
\caption{Errors with respect to FCI ($\Delta E$, in m\Eh), number of variational parameters ($N_{\rm par}$), and non-parallelism error (NPE = $| \Delta E(r=3) - \Delta  E(r=1.1)|$) for the ground state of N$_{2}$ at $r$ = 1.1 and 3 \AA{} computed with the cc-pVDZ basis set.  ACI and MR-CI computations used restricted Hartree--Fock and CASSCF(6,6) orbitals, respectively.  The 1s-like orbitals of nitrogen were frozen in all correlated computations. For $\sigma=1$, we also report ACI results computed using MP2 natural orbitals (NO) and CASSCF(6,6) orbitals (CAS).}
\footnotesize
\ifpreprint
	\begin{tabular*}{0.75\textwidth}{l @{\extracolsep{\fill}} rrrrr}
\else
	\begin{tabular*}{0.5\textwidth}{l @{\extracolsep{\fill}} rrrrr}
\fi
	\hline
	
	\hline
	           & \multicolumn{2}{c}{$r = 1.1$ \AA{}} & \multicolumn{2}{c}{$r = 3$ \AA{}} & NPE \\ \cline{2-3} \cline {4-5}
	           & $\Delta E$ & $N_{\rm par}$$^a$ & $\Delta E$  & $N_{\rm par}$ \\ \hline
    MR-CISD    &  9.02 & 5352/28030 &  8.14  & 5352/28030  & 0.88 \\
    MR-CISD+Q  & $-0.83$ & 5352/28030 & $-1.96$  &  5352/28030   & 1.13 \\
    \\
    ACI(50)  & 50.73 & 963    &  54.02 &  8044     & 3.29 \\
    ACI(10)  & 11.20 & 23940  &  10.88 &  54008    & 0.32  \\
    ACI(5) & 5.00  & 104398 &   4.59  &  308804   & 0.41  \\
    ACI(1) & 0.91  & 613198 &  0.78  &  1727993  & 0.13  \\
    \\
    ACI(1) (CAS) & 0.90  & 369562 &  0.69  &  1338097 & 0.11  \\
    ACI(1) (NO)   & 0.87  & 348789 &  0.78  &  1494181  & 0.09 \\
    \\
    ACI(50)+PT2  & 0.73    & 963    &   4.03  &  8044     & 3.30 \\
    ACI(10)+PT2  & 1.20    & 23940  &   0.88  &  54008    & 0.32  \\
    ACI(5)+PT2 & $-0.01$ & 104398 & $-0.41$ &  308804   & 0.40  \\
    ACI(1)+PT2 & $-0.09$  & 613198 & $-0.22$ &  1727993  & 0.13  \\
    \\
    FCI                 &    & 540924024 & & 540924024 & \\
    \hline
    
    \hline
    \end{tabular*}
    \\
        $^a$ For MR-CISD and MR-CISD+Q we report the total number of contracted/uncontracted configuration state functions.
    \label{tab:n2_tab}
\end{table}

Table \ref{tab:n2_tab} compares the energy error with respect to FCI ($\Delta E$) and the size of the ACI determinant space for N$_2$ at the bond distances 1.1 and 3 \AA{}.
In both cases, ACI energy errors with respect to FCI show very good correlation with the value of $\sigma$.
For a given value of $\sigma$, the energy difference $| \Delta E(r=3) - \Delta  E(r=1.1)|$ is only a fraction of the absolute error, showing the ability of the ACI method to describe both static and dynamic correlation in a balanced way.
With the perturbative correction, the absolute energy errors are further reduced but the NPEs remain virtually unchanged.
When we use natural orbitals from second-order M{\o}ller--Plesset perturbation theory or CASSCF (see Table \ref{tab:n2_tab}), the ACI(1) gives a more compact model space, with improved energy error with respect to RHF orbitals.
This result suggests that the parameter $\sigma$ effectively controls the ACI error regardless of the molecular orbital basis.

\begin{table*}[ht!]
    \footnotesize    \caption{Singlet-triplet splitting of the acene series computed with the ACI , DMRG, and v-2RDM methods using the STO-3G basis set.
    All carbon $\pi$ orbitals were correlated.}
    \begin{tabular*}{\textwidth}{c @{\extracolsep{\fill}} crrrrrrccccc}
    \hline
    
    \hline
    & \multicolumn{4}{c}{Singlet} & \multicolumn{3}{c}{Triplet} & \multicolumn{5}{c}{$\Delta E_{\rm{ST}}$ (\kcal)} \\ \cline{3-5} \cline{6-8} \cline{9-13}
    $n$ & CAS      & ACI(100) & ACI(50) &  ACI(10) & ACI(100) & ACI(50) & ACI(10) & ACI(100) & ACI(50) & ACI(10) & v-2RDM & DMRG \\ \hline
    2 & (10,10)  & 14       & 41      & 332         & 16    & 43       &  376      & 67.4 & 66.3 & 62.4 & 63.8 & 61.5 \\
    3 & (14,14)  & 76       & 230     & 8325        & 73    & 249      & 8600      & 51.8 & 51.0 & 46.8 & 45.2 & 46.0 \\
    4 & (18,18)  & 278      & 930     & 136190      & 280   & 1097     & 146814    & 38.4 & 38.3 & 35.5 & 32.8 & 34.7 \\ 
    5 & (22,22)  & 821      & 3444    & 1260702     & 849   & 4990     & 1495276   & 30.5 & 29.4 & 27.4 & 24.5 & 26.7 \\  
    6 & (26,26)  & 2174     & 31294   & 2770391     & 2220  & 40774    & 3352196   & 25.3 & 22.4 & 21.0 & 19.7 & 21.0 \\  
    8 & (34,34)  & 10580    & 1677179 &             & 11140 & 1496690  &           & 13.0 & 15.3 &      & 15.4 & 14.2 \\  
   10 & (42,42)  & 82403    &         &             & 81034 &          &           & 9.1  &      &      & 13.0 & 11.6 \\
   \hline
   
   \hline
    \end{tabular*}
    \label{tab:acenes}
\end{table*}

%To demonstrate the ability of ACI in accurately exploiting the sparsity of CASCI spaces, we turn to the polyacene series.
%Polyacenes and graphene patches have been of interest due to the high delocalization of electrons involved, and the interesting electronic properties that result.\cite{kivelson1983polyacene,raghu2002structural}
%Despite some debate over the ground state multiplicity of polyacenes, theoretical and experimental evidence now generally support the claim of a ground state singlet, though the open-shell character of the singlet ground state is not entirely resolved.\cite{hachmann2007radical,hajgato2009benchmark,hajgato2011focal,mizukami2012more,rivero2013entanglement, ibeji2015singlet, fosso2015accuracy}

To demonstrate the ability of ACI in accurately exploiting the sparsity of CASCI spaces, we turn to the polyacene series.\cite{hachmann2007radical,hajgato2009benchmark,hajgato2011focal,mizukami2012more,rivero2013entanglement, ibeji2015singlet, fosso2015accuracy}
The polyacene geometries from Ref.~\citenum{hachmann2007radical} were used in our calculations, and only the $\pi$ bonding and antibonding pairs included in the STO-3G basis were correlated.
This corresponds to a CAS($4n+2$,$4n+2$) wave function, where $n$ is the number of fused benzene rings.
Such a problem is intractable with CASCI for $n \geq 4$, but it is well suited for DMRG\cite{hachmann2007radical}
and the two-electron reduced density matrix (v-2RDM) method.\cite{gidofalvi2008active,mazziotti2011large, fosso2015accuracy}
%, which currently has described up to dodecacene in a CAS(50,50).
To directly compare with previous results,\cite{hachmann2007radical,fosso2015accuracy} all ACI computations use canonical RHF orbitals.

Table \ref{tab:acenes} shows the vertical singlet-triplet splittings ($\Delta E_{\rm ST} = E^{S=1} - E^{S=0}$) and the required number of determinants for various values of $\sigma$, in addition
to comparison with DMRG\cite{hachmann2007radical} and v-2RDM\cite{fosso2015accuracy} results. Note that to guarantee sub-\kcal accuracy, 
a $\sigma$ value less than $~1.6$ m\Eh is in principle required. However, already for $\sigma = 10$ m\Eh, the ACI error with respect to DMRG is consistently less that 1 \kcal through hexacene. For
$\sigma = 50$ m\Eh, we see the maximum error at anthracene, and in general the errors in the singlet-triplet splitting decrease with increasing $n$. A similar
trend is seen with the v-2RDM data, where the maximum in error is at $n=5$.
Our current pilot ACI code can be used to perform computations with up to about $5 \times 10^6$ determinants, which currently limits the ACI(10) to hexacene and the ACI(100) to decacene.  However, we anticipate that a production-level implementation of the ACI method that can take advantage of distributed memory architectures will be able to routinely target $10^7$--$10^8$ determinants. \cite{stampfuss2005improved}
Furthermore, we anticipate that like in the case of the DMRG,\cite{chan2002highly,moritz2005convergence} in the context of the ACI a localized molecular orbital basis will be crucial to significantly compress the number of variational parameters, and in turn, expand the applicability of this method to larger active spaces.

In summary, the major benefits of the ACI method are that: i) electron correlation can be treated in a balanced way without \textit{a priori} knowledge of a system's electronic structure and ii) that the energy error is precisely controlled by one user-specified parameter.
In addition, the ACI can be easily extended to excited states and implemented on distributed memory architectures.
The most practical use of ACI is as a reference wave function in multireference perturbative and non-perturbative treatments of electron correlation.
The straightforward computation of ACI reduced density matrices enables this extension.
Therefore, the ACI is an interesting alternative to DMRG, MCCI, FCIQMC, and v-2RDM methods.

All ACI results were obtained using our pilot code (\textsc{Forte}),\cite{FORTE2015} which is a suite of multireference methods written as a plugin to open-source quantum chemistry package \textsc{Psi4}.\cite{turney2012psi4}
This work was supported by start-up funds provided by Emory University.
\bibliography{arxiv_aci}

\begin{thebibliography}{49}%
\makeatletter
\providecommand \@ifxundefined [1]{%
 \@ifx{#1\undefined}
}%
\providecommand \@ifnum [1]{%
 \ifnum #1\expandafter \@firstoftwo
 \else \expandafter \@secondoftwo
 \fi
}%
\providecommand \@ifx [1]{%
 \ifx #1\expandafter \@firstoftwo
 \else \expandafter \@secondoftwo
 \fi
}%
\providecommand \natexlab [1]{#1}%
\providecommand \enquote  [1]{``#1''}%
\providecommand \bibnamefont  [1]{#1}%
\providecommand \bibfnamefont [1]{#1}%
\providecommand \citenamefont [1]{#1}%
\providecommand \href@noop [0]{\@secondoftwo}%
\providecommand \href [0]{\begingroup \@sanitize@url \@href}%
\providecommand \@href[1]{\@@startlink{#1}\@@href}%
\providecommand \@@href[1]{\endgroup#1\@@endlink}%
\providecommand \@sanitize@url [0]{\catcode `\\12\catcode `\$12\catcode
  `\&12\catcode `\#12\catcode `\^12\catcode `\_12\catcode `\%12\relax}%
\providecommand \@@startlink[1]{}%
\providecommand \@@endlink[0]{}%
\providecommand \url  [0]{\begingroup\@sanitize@url \@url }%
\providecommand \@url [1]{\endgroup\@href {#1}{\urlprefix }}%
\providecommand \urlprefix  [0]{URL }%
\providecommand \Eprint [0]{\href }%
\providecommand \doibase [0]{http://dx.doi.org/}%
\providecommand \selectlanguage [0]{\@gobble}%
\providecommand \bibinfo  [0]{\@secondoftwo}%
\providecommand \bibfield  [0]{\@secondoftwo}%
\providecommand \translation [1]{[#1]}%
\providecommand \BibitemOpen [0]{}%
\providecommand \bibitemStop [0]{}%
\providecommand \bibitemNoStop [0]{.\EOS\space}%
\providecommand \EOS [0]{\spacefactor3000\relax}%
\providecommand \BibitemShut  [1]{\csname bibitem#1\endcsname}%
\let\auto@bib@innerbib\@empty
%</preamble>
\bibitem [{\citenamefont {Sherrill}\ and\ \citenamefont
  {Schaefer~III}(1999)}]{sherrillCIrev}%
  \BibitemOpen
  \bibfield  {author} {\bibinfo {author} {\bibfnamefont {C.~D.}\ \bibnamefont
  {Sherrill}}\ and\ \bibinfo {author} {\bibfnamefont {H.~F.}\ \bibnamefont
  {Schaefer~III}},\ }\href@noop {} {\bibfield  {journal} {\bibinfo  {journal}
  {Adv. Quant. Chem.}\ }\textbf {\bibinfo {volume} {34}},\ \bibinfo {pages}
  {143} (\bibinfo {year} {1999})}\BibitemShut {NoStop}%
\bibitem [{\citenamefont {White}(1992)}]{white1992density}%
  \BibitemOpen
  \bibfield  {author} {\bibinfo {author} {\bibfnamefont {S.~R.}\ \bibnamefont
  {White}},\ }\href@noop {} {\bibfield  {journal} {\bibinfo  {journal} {Phys.
  Rev. Lett.}\ }\textbf {\bibinfo {volume} {69}},\ \bibinfo {pages} {2863}
  (\bibinfo {year} {1992})}\BibitemShut {NoStop}%
\bibitem [{\citenamefont {Chan}\ and\ \citenamefont
  {Head-Gordon}(2002)}]{chan2002highly}%
  \BibitemOpen
  \bibfield  {author} {\bibinfo {author} {\bibfnamefont {G.~K.-L.}\
  \bibnamefont {Chan}}\ and\ \bibinfo {author} {\bibfnamefont {M.}~\bibnamefont
  {Head-Gordon}},\ }\href {\doibase http://dx.doi.org/10.1063/1.1449459}
  {\bibfield  {journal} {\bibinfo  {journal} {J. Chem. Phys.}\ }\textbf
  {\bibinfo {volume} {116}},\ \bibinfo {pages} {4462} (\bibinfo {year}
  {2002})}\BibitemShut {NoStop}%
\bibitem [{\citenamefont {Moritz}, \citenamefont {Wolf},\ and\ \citenamefont
  {Reiher}(2005)}]{moritz2005relativistic}%
  \BibitemOpen
  \bibfield  {author} {\bibinfo {author} {\bibfnamefont {G.}~\bibnamefont
  {Moritz}}, \bibinfo {author} {\bibfnamefont {A.}~\bibnamefont {Wolf}}, \ and\
  \bibinfo {author} {\bibfnamefont {M.}~\bibnamefont {Reiher}},\ }\href@noop {}
  {\bibfield  {journal} {\bibinfo  {journal} {J. Chem. Phys.}\ }\textbf
  {\bibinfo {volume} {123}},\ \bibinfo {pages} {184105} (\bibinfo {year}
  {2005})}\BibitemShut {NoStop}%
\bibitem [{\citenamefont {Kurashige}\ and\ \citenamefont
  {Yanai}(2009)}]{kurashige2009high}%
  \BibitemOpen
  \bibfield  {author} {\bibinfo {author} {\bibfnamefont {Y.}~\bibnamefont
  {Kurashige}}\ and\ \bibinfo {author} {\bibfnamefont {T.}~\bibnamefont
  {Yanai}},\ }\href@noop {} {\bibfield  {journal} {\bibinfo  {journal} {J.
  Chem. Phys.}\ }\textbf {\bibinfo {volume} {130}},\ \bibinfo {pages} {234114}
  (\bibinfo {year} {2009})}\BibitemShut {NoStop}%
\bibitem [{\citenamefont {Olivares-Amaya}\ \emph {et~al.}(2015)\citenamefont
  {Olivares-Amaya}, \citenamefont {Hu}, \citenamefont {Nakatani}, \citenamefont
  {Sharma}, \citenamefont {Yang},\ and\ \citenamefont {Chan}}]{olivares2015ab}%
  \BibitemOpen
  \bibfield  {author} {\bibinfo {author} {\bibfnamefont {R.}~\bibnamefont
  {Olivares-Amaya}}, \bibinfo {author} {\bibfnamefont {W.}~\bibnamefont {Hu}},
  \bibinfo {author} {\bibfnamefont {N.}~\bibnamefont {Nakatani}}, \bibinfo
  {author} {\bibfnamefont {S.}~\bibnamefont {Sharma}}, \bibinfo {author}
  {\bibfnamefont {J.}~\bibnamefont {Yang}}, \ and\ \bibinfo {author}
  {\bibfnamefont {G.~K.-L.}\ \bibnamefont {Chan}},\ }\href@noop {} {\bibfield
  {journal} {\bibinfo  {journal} {J. Chem. Phys.}\ }\textbf {\bibinfo {volume}
  {142}},\ \bibinfo {pages} {034102} (\bibinfo {year} {2015})}\BibitemShut
  {NoStop}%
\bibitem [{\citenamefont {Greer}(1995)}]{greer1995estimating}%
  \BibitemOpen
  \bibfield  {author} {\bibinfo {author} {\bibfnamefont {J.}~\bibnamefont
  {Greer}},\ }\href@noop {} {\bibfield  {journal} {\bibinfo  {journal} {J.
  Chem. Phys.}\ }\textbf {\bibinfo {volume} {103}},\ \bibinfo {pages} {1821}
  (\bibinfo {year} {1995})}\BibitemShut {NoStop}%
\bibitem [{\citenamefont {Kelly}\ \emph {et~al.}(2014)\citenamefont {Kelly},
  \citenamefont {Perera}, \citenamefont {Bartlett},\ and\ \citenamefont
  {Greer}}]{kelly2014monte}%
  \BibitemOpen
  \bibfield  {author} {\bibinfo {author} {\bibfnamefont {T.~P.}\ \bibnamefont
  {Kelly}}, \bibinfo {author} {\bibfnamefont {A.}~\bibnamefont {Perera}},
  \bibinfo {author} {\bibfnamefont {R.~J.}\ \bibnamefont {Bartlett}}, \ and\
  \bibinfo {author} {\bibfnamefont {J.~C.}\ \bibnamefont {Greer}},\ }\href@noop
  {} {\bibfield  {journal} {\bibinfo  {journal} {J. Chem. Phys.}\ }\textbf
  {\bibinfo {volume} {140}},\ \bibinfo {pages} {084114} (\bibinfo {year}
  {2014})}\BibitemShut {NoStop}%
\bibitem [{\citenamefont {Coe}\ and\ \citenamefont
  {Paterson}(2012)}]{coe2012development}%
  \BibitemOpen
  \bibfield  {author} {\bibinfo {author} {\bibfnamefont {J.}~\bibnamefont
  {Coe}}\ and\ \bibinfo {author} {\bibfnamefont {M.}~\bibnamefont {Paterson}},\
  }\href@noop {} {\bibfield  {journal} {\bibinfo  {journal} {J. Chem. Phys.}\
  }\textbf {\bibinfo {volume} {137}},\ \bibinfo {pages} {204108} (\bibinfo
  {year} {2012})}\BibitemShut {NoStop}%
\bibitem [{\citenamefont {Coe}, \citenamefont {Murphy},\ and\ \citenamefont
  {Paterson}(2014)}]{coe2014applying}%
  \BibitemOpen
  \bibfield  {author} {\bibinfo {author} {\bibfnamefont {J.}~\bibnamefont
  {Coe}}, \bibinfo {author} {\bibfnamefont {P.}~\bibnamefont {Murphy}}, \ and\
  \bibinfo {author} {\bibfnamefont {M.}~\bibnamefont {Paterson}},\ }\href@noop
  {} {\bibfield  {journal} {\bibinfo  {journal} {Chem. Phys. Lett.}\ }\textbf
  {\bibinfo {volume} {604}},\ \bibinfo {pages} {46} (\bibinfo {year}
  {2014})}\BibitemShut {NoStop}%
\bibitem [{\citenamefont {Booth}, \citenamefont {Thom},\ and\ \citenamefont
  {Alavi}(2009)}]{booth2009fermion}%
  \BibitemOpen
  \bibfield  {author} {\bibinfo {author} {\bibfnamefont {G.~H.}\ \bibnamefont
  {Booth}}, \bibinfo {author} {\bibfnamefont {A.~J.}\ \bibnamefont {Thom}}, \
  and\ \bibinfo {author} {\bibfnamefont {A.}~\bibnamefont {Alavi}},\
  }\href@noop {} {\bibfield  {journal} {\bibinfo  {journal} {J. Chem. Phys.}\
  }\textbf {\bibinfo {volume} {131}},\ \bibinfo {pages} {054106} (\bibinfo
  {year} {2009})}\BibitemShut {NoStop}%
\bibitem [{\citenamefont {Cleland}, \citenamefont {Booth},\ and\ \citenamefont
  {Alavi}(2010)}]{cleland2010communications}%
  \BibitemOpen
  \bibfield  {author} {\bibinfo {author} {\bibfnamefont {D.}~\bibnamefont
  {Cleland}}, \bibinfo {author} {\bibfnamefont {G.~H.}\ \bibnamefont {Booth}},
  \ and\ \bibinfo {author} {\bibfnamefont {A.}~\bibnamefont {Alavi}},\
  }\href@noop {} {\bibfield  {journal} {\bibinfo  {journal} {J. Chem. Phys.}\
  }\textbf {\bibinfo {volume} {132}},\ \bibinfo {pages} {041103} (\bibinfo
  {year} {2010})}\BibitemShut {NoStop}%
\bibitem [{\citenamefont {Petruzielo}\ \emph {et~al.}(2012)\citenamefont
  {Petruzielo}, \citenamefont {Holmes}, \citenamefont {Changlani},
  \citenamefont {Nightingale},\ and\ \citenamefont
  {Umrigar}}]{Petruzielo:2012ha}%
  \BibitemOpen
  \bibfield  {author} {\bibinfo {author} {\bibfnamefont {F.~R.}\ \bibnamefont
  {Petruzielo}}, \bibinfo {author} {\bibfnamefont {A.~A.}\ \bibnamefont
  {Holmes}}, \bibinfo {author} {\bibfnamefont {H.~J.}\ \bibnamefont
  {Changlani}}, \bibinfo {author} {\bibfnamefont {M.~P.}\ \bibnamefont
  {Nightingale}}, \ and\ \bibinfo {author} {\bibfnamefont {C.~J.}\ \bibnamefont
  {Umrigar}},\ }\href@noop {} {\bibfield  {journal} {\bibinfo  {journal} {Phys.
  Rev. Lett.}\ }\textbf {\bibinfo {volume} {109}},\ \bibinfo {pages} {230201}
  (\bibinfo {year} {2012})}\BibitemShut {NoStop}%
\bibitem [{\citenamefont {Ten-no}(2013)}]{Tenno:2013kd}%
  \BibitemOpen
  \bibfield  {author} {\bibinfo {author} {\bibfnamefont {S.}~\bibnamefont
  {Ten-no}},\ }\href@noop {} {\bibfield  {journal} {\bibinfo  {journal} {J.
  Chem. Phys.}\ }\textbf {\bibinfo {volume} {138}},\ \bibinfo {pages} {164126}
  (\bibinfo {year} {2013})}\BibitemShut {NoStop}%
\bibitem [{\citenamefont {Kurashige}, \citenamefont {Chan},\ and\ \citenamefont
  {Yanai}(2013)}]{kurashige2013entangled}%
  \BibitemOpen
  \bibfield  {author} {\bibinfo {author} {\bibfnamefont {Y.}~\bibnamefont
  {Kurashige}}, \bibinfo {author} {\bibfnamefont {G.~K.-L.}\ \bibnamefont
  {Chan}}, \ and\ \bibinfo {author} {\bibfnamefont {T.}~\bibnamefont {Yanai}},\
  }\href@noop {} {\bibfield  {journal} {\bibinfo  {journal} {Nature Chem.}\
  }\textbf {\bibinfo {volume} {5}},\ \bibinfo {pages} {660} (\bibinfo {year}
  {2013})}\BibitemShut {NoStop}%
\bibitem [{\citenamefont {Booth}\ \emph {et~al.}(2013)\citenamefont {Booth},
  \citenamefont {Gr{\"u}neis}, \citenamefont {Kresse},\ and\ \citenamefont
  {Alavi}}]{Booth:2013er}%
  \BibitemOpen
  \bibfield  {author} {\bibinfo {author} {\bibfnamefont {G.~H.}\ \bibnamefont
  {Booth}}, \bibinfo {author} {\bibfnamefont {A.}~\bibnamefont {Gr{\"u}neis}},
  \bibinfo {author} {\bibfnamefont {G.}~\bibnamefont {Kresse}}, \ and\ \bibinfo
  {author} {\bibfnamefont {A.}~\bibnamefont {Alavi}},\ }\href@noop {}
  {\bibfield  {journal} {\bibinfo  {journal} {Nature}\ }\textbf {\bibinfo
  {volume} {493}},\ \bibinfo {pages} {365} (\bibinfo {year}
  {2013})}\BibitemShut {NoStop}%
\bibitem [{\citenamefont {Bender}\ and\ \citenamefont
  {Davidson}(1969)}]{bender1969studies}%
  \BibitemOpen
  \bibfield  {author} {\bibinfo {author} {\bibfnamefont {C.~F.}\ \bibnamefont
  {Bender}}\ and\ \bibinfo {author} {\bibfnamefont {E.~R.}\ \bibnamefont
  {Davidson}},\ }\href@noop {} {\bibfield  {journal} {\bibinfo  {journal}
  {Phys. Rev.}\ }\textbf {\bibinfo {volume} {183}},\ \bibinfo {pages} {23}
  (\bibinfo {year} {1969})}\BibitemShut {NoStop}%
\bibitem [{\citenamefont {Huron}, \citenamefont {Malrieu},\ and\ \citenamefont
  {Rancurel}(1973)}]{huron1973iterative}%
  \BibitemOpen
  \bibfield  {author} {\bibinfo {author} {\bibfnamefont {B.}~\bibnamefont
  {Huron}}, \bibinfo {author} {\bibfnamefont {J.~P.}\ \bibnamefont {Malrieu}},
  \ and\ \bibinfo {author} {\bibfnamefont {P.}~\bibnamefont {Rancurel}},\
  }\href@noop {} {\bibfield  {journal} {\bibinfo  {journal} {J. Chem. Phys.}\
  }\textbf {\bibinfo {volume} {58}},\ \bibinfo {pages} {5745} (\bibinfo {year}
  {1973})}\BibitemShut {NoStop}%
\bibitem [{\citenamefont {Buenker}\ and\ \citenamefont
  {Peyerimhoff}(1974)}]{buenker1974individualized}%
  \BibitemOpen
  \bibfield  {author} {\bibinfo {author} {\bibfnamefont {R.~J.}\ \bibnamefont
  {Buenker}}\ and\ \bibinfo {author} {\bibfnamefont {S.~D.}\ \bibnamefont
  {Peyerimhoff}},\ }\href@noop {} {\bibfield  {journal} {\bibinfo  {journal}
  {Theor. Chim. Acta}\ }\textbf {\bibinfo {volume} {35}},\ \bibinfo {pages}
  {33} (\bibinfo {year} {1974})}\BibitemShut {NoStop}%
\bibitem [{\citenamefont {Harrison}(1991)}]{harrison1991approximating}%
  \BibitemOpen
  \bibfield  {author} {\bibinfo {author} {\bibfnamefont {R.~J.}\ \bibnamefont
  {Harrison}},\ }\href@noop {} {\bibfield  {journal} {\bibinfo  {journal} {J.
  Chem. Phys.}\ }\textbf {\bibinfo {volume} {94}},\ \bibinfo {pages} {5021}
  (\bibinfo {year} {1991})}\BibitemShut {NoStop}%
\bibitem [{\citenamefont {Garc{\'\i}a}\ \emph {et~al.}(1995)\citenamefont
  {Garc{\'\i}a}, \citenamefont {Castell}, \citenamefont {Caballol},\ and\
  \citenamefont {Malrieu}}]{garcia1995iterative}%
  \BibitemOpen
  \bibfield  {author} {\bibinfo {author} {\bibfnamefont {V.}~\bibnamefont
  {Garc{\'\i}a}}, \bibinfo {author} {\bibfnamefont {O.}~\bibnamefont
  {Castell}}, \bibinfo {author} {\bibfnamefont {R.}~\bibnamefont {Caballol}}, \
  and\ \bibinfo {author} {\bibfnamefont {J.}~\bibnamefont {Malrieu}},\
  }\href@noop {} {\bibfield  {journal} {\bibinfo  {journal} {Chem. Phys.
  Lett.}\ }\textbf {\bibinfo {volume} {238}},\ \bibinfo {pages} {222} (\bibinfo
  {year} {1995})}\BibitemShut {NoStop}%
\bibitem [{\citenamefont {Neese}(2003)}]{neese2003spectroscopy}%
  \BibitemOpen
  \bibfield  {author} {\bibinfo {author} {\bibfnamefont {F.}~\bibnamefont
  {Neese}},\ }\href@noop {} {\bibfield  {journal} {\bibinfo  {journal} {J.
  Chem. Phys.}\ }\textbf {\bibinfo {volume} {119}},\ \bibinfo {pages} {9428}
  (\bibinfo {year} {2003})}\BibitemShut {NoStop}%
\bibitem [{\citenamefont {Nakatsuji}\ and\ \citenamefont
  {Ehara}(2005)}]{nakatsuji2005iterative}%
  \BibitemOpen
  \bibfield  {author} {\bibinfo {author} {\bibfnamefont {H.}~\bibnamefont
  {Nakatsuji}}\ and\ \bibinfo {author} {\bibfnamefont {M.}~\bibnamefont
  {Ehara}},\ }\href {\doibase http://dx.doi.org/10.1063/1.1898207} {\bibfield
  {journal} {\bibinfo  {journal} {J. Chem. Phys.}\ }\textbf {\bibinfo {volume}
  {122}},\ \bibinfo {pages} {194108} (\bibinfo {year} {2005})}\BibitemShut
  {NoStop}%
\bibitem [{\citenamefont {Abrams}\ and\ \citenamefont
  {Sherrill}(2005)}]{abrams2005important}%
  \BibitemOpen
  \bibfield  {author} {\bibinfo {author} {\bibfnamefont {M.~L.}\ \bibnamefont
  {Abrams}}\ and\ \bibinfo {author} {\bibfnamefont {C.~D.}\ \bibnamefont
  {Sherrill}},\ }\href@noop {} {\bibfield  {journal} {\bibinfo  {journal}
  {Chem. Phys. Lett.}\ }\textbf {\bibinfo {volume} {412}},\ \bibinfo {pages}
  {121} (\bibinfo {year} {2005})}\BibitemShut {NoStop}%
\bibitem [{\citenamefont {Bytautas}\ and\ \citenamefont
  {Ruedenberg}(2009)}]{bytautas2009priori}%
  \BibitemOpen
  \bibfield  {author} {\bibinfo {author} {\bibfnamefont {L.}~\bibnamefont
  {Bytautas}}\ and\ \bibinfo {author} {\bibfnamefont {K.}~\bibnamefont
  {Ruedenberg}},\ }\href@noop {} {\bibfield  {journal} {\bibinfo  {journal}
  {Chem. Phys.}\ }\textbf {\bibinfo {volume} {356}},\ \bibinfo {pages} {64}
  (\bibinfo {year} {2009})}\BibitemShut {NoStop}%
\bibitem [{\citenamefont {Roth}(2009)}]{roth2009importance}%
  \BibitemOpen
  \bibfield  {author} {\bibinfo {author} {\bibfnamefont {R.}~\bibnamefont
  {Roth}},\ }\href@noop {} {\bibfield  {journal} {\bibinfo  {journal} {Phys.
  Rev. C}\ }\textbf {\bibinfo {volume} {79}},\ \bibinfo {pages} {064324}
  (\bibinfo {year} {2009})}\BibitemShut {NoStop}%
\bibitem [{\citenamefont {Evangelista}(2014)}]{evangelista2014adaptive}%
  \BibitemOpen
  \bibfield  {author} {\bibinfo {author} {\bibfnamefont {F.~A.}\ \bibnamefont
  {Evangelista}},\ }\href {\doibase http://dx.doi.org/10.1063/1.4869192}
  {\bibfield  {journal} {\bibinfo  {journal} {J. Chem. Phys.}\ }\textbf
  {\bibinfo {volume} {140}},\ \bibinfo {pages} {054109} (\bibinfo {year}
  {2014})}\BibitemShut {NoStop}%
\bibitem [{\citenamefont {Knowles}(2015)}]{knowles2015compressive}%
  \BibitemOpen
  \bibfield  {author} {\bibinfo {author} {\bibfnamefont {P.~J.}\ \bibnamefont
  {Knowles}},\ }\href@noop {} {\bibfield  {journal} {\bibinfo  {journal} {Mol.
  Phys.}\ }\textbf {\bibinfo {volume} {113}},\ \bibinfo {pages} {1655}
  (\bibinfo {year} {2015})}\BibitemShut {NoStop}%
\bibitem [{\citenamefont {Liu}\ and\ \citenamefont
  {Hoffmann}(2016)}]{liu2016ici}%
  \BibitemOpen
  \bibfield  {author} {\bibinfo {author} {\bibfnamefont {W.}~\bibnamefont
  {Liu}}\ and\ \bibinfo {author} {\bibfnamefont {M.~R.}\ \bibnamefont
  {Hoffmann}},\ }\href@noop {} {\bibfield  {journal} {\bibinfo  {journal} {J.
  Chem. Theory Comput.}\ } (\bibinfo {year} {2016})}\BibitemShut {NoStop}%
\bibitem [{\citenamefont {Tubman}\ \emph {et~al.}(2016)\citenamefont {Tubman},
  \citenamefont {Lee}, \citenamefont {Takeshita}, \citenamefont {Head-Gordon},\
  and\ \citenamefont {Whaley}}]{Tubman:2016wi}%
  \BibitemOpen
  \bibfield  {author} {\bibinfo {author} {\bibfnamefont {N.~M.}\ \bibnamefont
  {Tubman}}, \bibinfo {author} {\bibfnamefont {J.}~\bibnamefont {Lee}},
  \bibinfo {author} {\bibfnamefont {T.~Y.}\ \bibnamefont {Takeshita}}, \bibinfo
  {author} {\bibfnamefont {M.}~\bibnamefont {Head-Gordon}}, \ and\ \bibinfo
  {author} {\bibfnamefont {K.~B.}\ \bibnamefont {Whaley}},\ }\href@noop {} {\
  (\bibinfo {year} {2016})},\ \Eprint {http://arxiv.org/abs/arXiv:1603.02686
  [cond-mat.str-el]} {arXiv:1603.02686 [cond-mat.str-el]} \BibitemShut
  {NoStop}%
\bibitem [{\citenamefont {Assfeld}, \citenamefont {Alml{\"o}f},\ and\
  \citenamefont {Truhlar}(1995)}]{assfeld1995degeneracy}%
  \BibitemOpen
  \bibfield  {author} {\bibinfo {author} {\bibfnamefont {X.}~\bibnamefont
  {Assfeld}}, \bibinfo {author} {\bibfnamefont {J.~E.}\ \bibnamefont
  {Alml{\"o}f}}, \ and\ \bibinfo {author} {\bibfnamefont {D.~G.}\ \bibnamefont
  {Truhlar}},\ }\href@noop {} {\bibfield  {journal} {\bibinfo  {journal} {Chem.
  Phys. Lett.}\ }\textbf {\bibinfo {volume} {241}},\ \bibinfo {pages} {438}
  (\bibinfo {year} {1995})}\BibitemShut {NoStop}%
\bibitem [{\citenamefont {Angeli}\ and\ \citenamefont
  {Persico}(1997)}]{angeli1997multireference2}%
  \BibitemOpen
  \bibfield  {author} {\bibinfo {author} {\bibfnamefont {C.}~\bibnamefont
  {Angeli}}\ and\ \bibinfo {author} {\bibfnamefont {M.}~\bibnamefont
  {Persico}},\ }\href {\doibase 10.1007/s002140050285} {\bibfield  {journal}
  {\bibinfo  {journal} {Theor. Chem. Acc.}\ }\textbf {\bibinfo {volume} {98}},\
  \bibinfo {pages} {117} (\bibinfo {year} {1997})}\BibitemShut {NoStop}%
\bibitem [{1()}]{1}%
  \BibitemOpen
  \href@noop {} {}\bibinfo {note} {To maximize efficiency, ACI works in the
  basis of Slater determinants rather than configuration state functions.
  Consequently, $P^{(k)}$ and $M^{(k)}$ may not form spin complete sets. To
  bypass this issue, in certain cases we have enforced spin completeness by
  appropriate augmenting $P^{(k)}$ and $M^{(k)}$. In practice, correcting for
  spin incompleteness is only necessary to describe near-degenerate states of
  different spin. Therefore, in this work this procedure is only applied to our
  N$_2$ computations to recover the correct asymptotic dissociation
  limit.}\BibitemShut {Stop}%
\bibitem [{\citenamefont {Werner}\ and\ \citenamefont
  {Knowles}(1988)}]{werner1988efficient}%
  \BibitemOpen
  \bibfield  {author} {\bibinfo {author} {\bibfnamefont {H.-J.}\ \bibnamefont
  {Werner}}\ and\ \bibinfo {author} {\bibfnamefont {P.~J.}\ \bibnamefont
  {Knowles}},\ }\href@noop {} {\bibfield  {journal} {\bibinfo  {journal} {J.
  Chem. Phys.}\ }\textbf {\bibinfo {volume} {89}},\ \bibinfo {pages} {5803}
  (\bibinfo {year} {1988})}\BibitemShut {NoStop}%
\bibitem [{\citenamefont {Langhoff}\ and\ \citenamefont
  {Davidson}(1974)}]{langhoff1974configuration}%
  \BibitemOpen
  \bibfield  {author} {\bibinfo {author} {\bibfnamefont {S.~R.}\ \bibnamefont
  {Langhoff}}\ and\ \bibinfo {author} {\bibfnamefont {E.~R.}\ \bibnamefont
  {Davidson}},\ }\href@noop {} {\bibfield  {journal} {\bibinfo  {journal} {Int.
  J. Quant. Chem.}\ }\textbf {\bibinfo {volume} {8}},\ \bibinfo {pages} {61}
  (\bibinfo {year} {1974})}\BibitemShut {NoStop}%
\bibitem [{\citenamefont {Li}\ and\ \citenamefont
  {Evangelista}(2016)}]{li2016towards}%
  \BibitemOpen
  \bibfield  {author} {\bibinfo {author} {\bibfnamefont {C.}~\bibnamefont
  {Li}}\ and\ \bibinfo {author} {\bibfnamefont {F.~A.}\ \bibnamefont
  {Evangelista}},\ }\href@noop {} {\bibfield  {journal} {\bibinfo  {journal}
  {arXiv preprint arXiv:1602.05667}\ } (\bibinfo {year} {2016})}\BibitemShut
  {NoStop}%
\bibitem [{\citenamefont {Hachmann}\ \emph {et~al.}(2007)\citenamefont
  {Hachmann}, \citenamefont {Dorando}, \citenamefont {Avil{\'e}s},\ and\
  \citenamefont {Chan}}]{hachmann2007radical}%
  \BibitemOpen
  \bibfield  {author} {\bibinfo {author} {\bibfnamefont {J.}~\bibnamefont
  {Hachmann}}, \bibinfo {author} {\bibfnamefont {J.~J.}\ \bibnamefont
  {Dorando}}, \bibinfo {author} {\bibfnamefont {M.}~\bibnamefont {Avil{\'e}s}},
  \ and\ \bibinfo {author} {\bibfnamefont {G.~K.-L.}\ \bibnamefont {Chan}},\
  }\href@noop {} {\bibfield  {journal} {\bibinfo  {journal} {J. Chem. Phys.}\
  }\textbf {\bibinfo {volume} {127}},\ \bibinfo {pages} {134309} (\bibinfo
  {year} {2007})}\BibitemShut {NoStop}%
\bibitem [{\citenamefont {Hajgat{\'o}}\ \emph {et~al.}(2009)\citenamefont
  {Hajgat{\'o}}, \citenamefont {Szieberth}, \citenamefont {Geerlings},
  \citenamefont {De~Proft},\ and\ \citenamefont
  {Deleuze}}]{hajgato2009benchmark}%
  \BibitemOpen
  \bibfield  {author} {\bibinfo {author} {\bibfnamefont {B.}~\bibnamefont
  {Hajgat{\'o}}}, \bibinfo {author} {\bibfnamefont {D.}~\bibnamefont
  {Szieberth}}, \bibinfo {author} {\bibfnamefont {P.}~\bibnamefont
  {Geerlings}}, \bibinfo {author} {\bibfnamefont {F.}~\bibnamefont {De~Proft}},
  \ and\ \bibinfo {author} {\bibfnamefont {M.}~\bibnamefont {Deleuze}},\
  }\href@noop {} {\bibfield  {journal} {\bibinfo  {journal} {J. Chem. Phys.}\
  }\textbf {\bibinfo {volume} {131}},\ \bibinfo {pages} {224321} (\bibinfo
  {year} {2009})}\BibitemShut {NoStop}%
\bibitem [{\citenamefont {Hajgat{\'o}}, \citenamefont {Huzak},\ and\
  \citenamefont {Deleuze}(2011)}]{hajgato2011focal}%
  \BibitemOpen
  \bibfield  {author} {\bibinfo {author} {\bibfnamefont {B.}~\bibnamefont
  {Hajgat{\'o}}}, \bibinfo {author} {\bibfnamefont {M.}~\bibnamefont {Huzak}},
  \ and\ \bibinfo {author} {\bibfnamefont {M.~S.}\ \bibnamefont {Deleuze}},\
  }\href@noop {} {\bibfield  {journal} {\bibinfo  {journal} {J. Phys. Chem. A}\
  }\textbf {\bibinfo {volume} {115}},\ \bibinfo {pages} {9282} (\bibinfo {year}
  {2011})}\BibitemShut {NoStop}%
\bibitem [{\citenamefont {Mizukami}, \citenamefont {Kurashige},\ and\
  \citenamefont {Yanai}(2012)}]{mizukami2012more}%
  \BibitemOpen
  \bibfield  {author} {\bibinfo {author} {\bibfnamefont {W.}~\bibnamefont
  {Mizukami}}, \bibinfo {author} {\bibfnamefont {Y.}~\bibnamefont {Kurashige}},
  \ and\ \bibinfo {author} {\bibfnamefont {T.}~\bibnamefont {Yanai}},\
  }\href@noop {} {\bibfield  {journal} {\bibinfo  {journal} {J. Chem. Theory
  Comput.}\ }\textbf {\bibinfo {volume} {9}},\ \bibinfo {pages} {401} (\bibinfo
  {year} {2012})}\BibitemShut {NoStop}%
\bibitem [{\citenamefont {Rivero}, \citenamefont {Jim{\'e}nez-Hoyos},\ and\
  \citenamefont {Scuseria}(2013)}]{rivero2013entanglement}%
  \BibitemOpen
  \bibfield  {author} {\bibinfo {author} {\bibfnamefont {P.}~\bibnamefont
  {Rivero}}, \bibinfo {author} {\bibfnamefont {C.~A.}\ \bibnamefont
  {Jim{\'e}nez-Hoyos}}, \ and\ \bibinfo {author} {\bibfnamefont {G.~E.}\
  \bibnamefont {Scuseria}},\ }\href@noop {} {\bibfield  {journal} {\bibinfo
  {journal} {J. Phys. Chem. B}\ }\textbf {\bibinfo {volume} {117}},\ \bibinfo
  {pages} {12750} (\bibinfo {year} {2013})}\BibitemShut {NoStop}%
\bibitem [{\citenamefont {Ibeji}\ and\ \citenamefont
  {Ghosh}(2015)}]{ibeji2015singlet}%
  \BibitemOpen
  \bibfield  {author} {\bibinfo {author} {\bibfnamefont {C.~U.}\ \bibnamefont
  {Ibeji}}\ and\ \bibinfo {author} {\bibfnamefont {D.}~\bibnamefont {Ghosh}},\
  }\href@noop {} {\bibfield  {journal} {\bibinfo  {journal} {Phys. Chem. Chem.
  Phys.}\ }\textbf {\bibinfo {volume} {17}},\ \bibinfo {pages} {9849} (\bibinfo
  {year} {2015})}\BibitemShut {NoStop}%
\bibitem [{\citenamefont {Fosso-Tande}, \citenamefont {Nascimento},\ and\
  \citenamefont {DePrince~III}(2015)}]{fosso2015accuracy}%
  \BibitemOpen
  \bibfield  {author} {\bibinfo {author} {\bibfnamefont {J.}~\bibnamefont
  {Fosso-Tande}}, \bibinfo {author} {\bibfnamefont {D.~R.}\ \bibnamefont
  {Nascimento}}, \ and\ \bibinfo {author} {\bibfnamefont {A.~E.}\ \bibnamefont
  {DePrince~III}},\ }\href@noop {} {\bibfield  {journal} {\bibinfo  {journal}
  {Mol. Phys.}\ }\textbf {\bibinfo {volume} {114}},\ \bibinfo {pages} {1}
  (\bibinfo {year} {2015})}\BibitemShut {NoStop}%
\bibitem [{\citenamefont {Gidofalvi}\ and\ \citenamefont
  {Mazziotti}(2008)}]{gidofalvi2008active}%
  \BibitemOpen
  \bibfield  {author} {\bibinfo {author} {\bibfnamefont {G.}~\bibnamefont
  {Gidofalvi}}\ and\ \bibinfo {author} {\bibfnamefont {D.~A.}\ \bibnamefont
  {Mazziotti}},\ }\href@noop {} {\bibfield  {journal} {\bibinfo  {journal} {J.
  Chem. Phys.}\ }\textbf {\bibinfo {volume} {129}},\ \bibinfo {pages} {134108}
  (\bibinfo {year} {2008})}\BibitemShut {NoStop}%
\bibitem [{\citenamefont {Mazziotti}(2011)}]{mazziotti2011large}%
  \BibitemOpen
  \bibfield  {author} {\bibinfo {author} {\bibfnamefont {D.~A.}\ \bibnamefont
  {Mazziotti}},\ }\href@noop {} {\bibfield  {journal} {\bibinfo  {journal}
  {Phys. Rev. Lett.}\ }\textbf {\bibinfo {volume} {106}},\ \bibinfo {pages}
  {083001} (\bibinfo {year} {2011})}\BibitemShut {NoStop}%
\bibitem [{\citenamefont {Stampfu{\ss}}\ and\ \citenamefont
  {Wenzel}(2005)}]{stampfuss2005improved}%
  \BibitemOpen
  \bibfield  {author} {\bibinfo {author} {\bibfnamefont {P.}~\bibnamefont
  {Stampfu{\ss}}}\ and\ \bibinfo {author} {\bibfnamefont {W.}~\bibnamefont
  {Wenzel}},\ }\href@noop {} {\bibfield  {journal} {\bibinfo  {journal} {J.
  Chem. Phys.}\ }\textbf {\bibinfo {volume} {122}},\ \bibinfo {pages} {024110}
  (\bibinfo {year} {2005})}\BibitemShut {NoStop}%
\bibitem [{\citenamefont {Moritz}, \citenamefont {Hess},\ and\ \citenamefont
  {Reiher}(2005)}]{moritz2005convergence}%
  \BibitemOpen
  \bibfield  {author} {\bibinfo {author} {\bibfnamefont {G.}~\bibnamefont
  {Moritz}}, \bibinfo {author} {\bibfnamefont {B.~A.}\ \bibnamefont {Hess}}, \
  and\ \bibinfo {author} {\bibfnamefont {M.}~\bibnamefont {Reiher}},\
  }\href@noop {} {\bibfield  {journal} {\bibinfo  {journal} {J. Chem. Phys.}\
  }\textbf {\bibinfo {volume} {122}},\ \bibinfo {pages} {024107} (\bibinfo
  {year} {2005})}\BibitemShut {NoStop}%
\bibitem [{FOR(2015)}]{FORTE2015}%
  \BibitemOpen
  \href@noop {} {}\bibinfo {howpublished} {Forte, a suite of quantum chemistry
  methods for strongly correlated electrons. For the current version, see
  \url{https://github.com/evangelistalab/forte}} (\bibinfo {year}
  {2015})\BibitemShut {NoStop}%
\bibitem [{\citenamefont {Turney}\ \emph {et~al.}(2012)\citenamefont {Turney},
  \citenamefont {Simmonett}, \citenamefont {Parrish}, \citenamefont
  {Hohenstein}, \citenamefont {Evangelista}, \citenamefont {Fermann},
  \citenamefont {Mintz}, \citenamefont {Burns}, \citenamefont {Wilke},
  \citenamefont {Abrams} \emph {et~al.}}]{turney2012psi4}%
  \BibitemOpen
  \bibfield  {author} {\bibinfo {author} {\bibfnamefont {J.~M.}\ \bibnamefont
  {Turney}}, \bibinfo {author} {\bibfnamefont {A.~C.}\ \bibnamefont
  {Simmonett}}, \bibinfo {author} {\bibfnamefont {R.~M.}\ \bibnamefont
  {Parrish}}, \bibinfo {author} {\bibfnamefont {E.~G.}\ \bibnamefont
  {Hohenstein}}, \bibinfo {author} {\bibfnamefont {F.~A.}\ \bibnamefont
  {Evangelista}}, \bibinfo {author} {\bibfnamefont {J.~T.}\ \bibnamefont
  {Fermann}}, \bibinfo {author} {\bibfnamefont {B.~J.}\ \bibnamefont {Mintz}},
  \bibinfo {author} {\bibfnamefont {L.~A.}\ \bibnamefont {Burns}}, \bibinfo
  {author} {\bibfnamefont {J.~J.}\ \bibnamefont {Wilke}}, \bibinfo {author}
  {\bibfnamefont {M.~L.}\ \bibnamefont {Abrams}},  \emph {et~al.},\ }\href@noop
  {} {\bibfield  {journal} {\bibinfo  {journal} {WIREs: Comp. Mol. Sci.}\
  }\textbf {\bibinfo {volume} {2}},\ \bibinfo {pages} {556} (\bibinfo {year}
  {2012})}\BibitemShut {NoStop}%
\end{thebibliography}%
\end{document}